\documentstyle[floats,twocolumn,aps,psfig,prl]{revtex}

\begin{document}
\draft

%***********    This is for two columns *******************************
\twocolumn[\hsize\textwidth\columnwidth\hsize\csname @twocolumnfalse\endcsname

\title{Conductance oscillations in mesoscopic rings: microscopic versus
       macroscopic picture}  
\author{R. H\"aussler$^1$, E. Scheer$^1$, H. B. Weber$^2$,
        and H. v. L\"ohneysen$^{1,2}$}
\address{
  $^1$Physikalisches Institut, Universit\"at Karlsruhe,
      D-76128 Karlsruhe, Germany\\
  $^2$Forschungszentrum Karlsruhe, Institut f\"ur Nanotechnologie,
      D-76021 Karlsruhe, Germany
}

\maketitle

\begin{abstract}
The phase of Aharonov-Bohm oscillations in mesoscopic metal rings 
in the presence of a magnetic field can be
modulated by application of a DC-bias current $I_{DC}$.
We address the question of how a variation of $I_{DC}$ and hence
of the microscopic phases of the electronic wave functions results in the
macroscopic phase of the conductance oscillations.
Whereas the first one can be varied continuously the latter has to be
quantized for a ring in two-wire configuration by virtue of the Onsager
symmetry relations.
We observe a correlation between a phase flip by $\pm \pi$ and the amplitude
of the oscillations.
\end{abstract}

\pacs{PACS numbers: 73.23.-b, 73.50.-h, 85.30.St}

% body of paper here
% ***********    This is for two columns *******************************
\vskip2pc]

\section{Introduction}

In mesoscopic rings
exposed to a perpendicular magnetic field $B$
the interference of partial waves of electrons propagating
phase-coherently in opposite directions leads to oscillations of the 
magnetoconductance with a fundamental period
$B_{per} = \phi_0/A$, where $\phi_0 = h/e$ is the flux quantum and
$A$ the area of the ring~\cite{webb}.
In rings with finite width of the arms the interference of waves within one arm
of the ring results in a modulation of the amplitude and the phase of the
oscillations~\cite{stone}:
\begin{equation}\label{eq1}
\Delta G = g(B,E) \, \cos \left( 2\pi \frac{B A}{\phi_0} + \varphi(B,E) \right)
\end{equation}
The oscillations of $\Delta G$ with respect to $B$ are usually termed
Aharonov-Bohm (AB) effect although strictly speaking the AB effect
in its original meaning refers to a phase shift of electron waves
by a vector potential only~\cite{aharonov}.
The amplitude $g$ and the phase $\varphi$ are sample specific as they depend on
the microscopic arrangement of the scattering centres in the ring.
Both quantities are random functions of the magnetic field $B$ and the
energy $E$ of the electrons.
The typical scales in $B$ and $E$ for a variation of $g$ and $\varphi$ are
the correlation field $B_c$ and the Thouless energy
$E_c \approx h D/L^2$, where $D$ is the diffusion constant and
$L$ the sample length.

In mesoscopic devices in general, the symmetry of the magnetoconductance upon
magnetic-field reversal depends on the sample geometry~\cite{buttiker}.
In a two-wire configuration the current and voltage leads branch outside
the phase-coherent region of the electrons and the conductance $G$
is symmetric with respect to the magnetic field $B$, $G(B) = G(-B)$.
This symmetry relation does not hold for a four-wire configuration with a
bifurcation of the voltage and current leads within the phase-coherent region.
This behaviour is based on the fundamental Onsager relations which are
a consequence of time-reversal symmetry.
Onsager succeeded in deriving general reciprocal relations from the principle
of microscopic reversibility~\cite{onsager}.
The application of the Onsager relations to the electrical conductance as a
macroscopic quantity was discussed by Casimir~\cite{casimir}.
The necessity to distinguish between mesoscopic two-wire and four-wire
configurations was demonstrated theoretically by B\"uttiker~\cite{buttiker}.

These symmetry relations have been confirmed experimentally in many
experiments, e.g. with mesoscopic rings~\cite{benoit}.
For a ring in two-wire configuration the symmetry condition does not
allow arbitrary values of the phase of the AB oscillations.
On the other hand, in a microscopic picture this phase is determined
by the arbitrary (but fixed for a given field) difference between electrons
travelling through the two arms which in turn depends on the phase
an electron accumulates at the scattering centres in the metal ring
during its diffusive motion.
In our experiment we address the question of how the
quantization condition of the macroscopically observable phase of the
conductance oscillations is fulfilled when the microscopic electronic phase
is varied continuously.
By application of a DC-bias current $I_{DC}$ we generate a 
non-equilibrium energy distribution of the electrons in the
ring since no energy relaxation occurs in the phase-coherent region.
The energy of the electrons contributing to the charge transport and hence
also the phase of the microscopic electronic wave function can be modified
by a variation of $I_{DC}$.
We analyze the reaction of the phase of the AB oscillations
on this continuous variation on the microscopic level by monitoring
the cross-correlation for magnetoconductance traces at different $I_{DC}$
on the macroscopic level.

\section{Experimental}

The samples were prepared by electron-beam lithography and evaporation of
Cu or Ag on a Si substrate followed by a lift-off process.
The differential resistance $dV/dI$ was measured at
19~Hz with a superimposed DC-bias current $I_{DC}$
with the sample mounted in the mixing chamber of a dilution
refrigerator with a base temperature of 20~mK.

\begin{figure}
   \begin{center}
      \begin{tabular}{ll} 
         \psfig{file=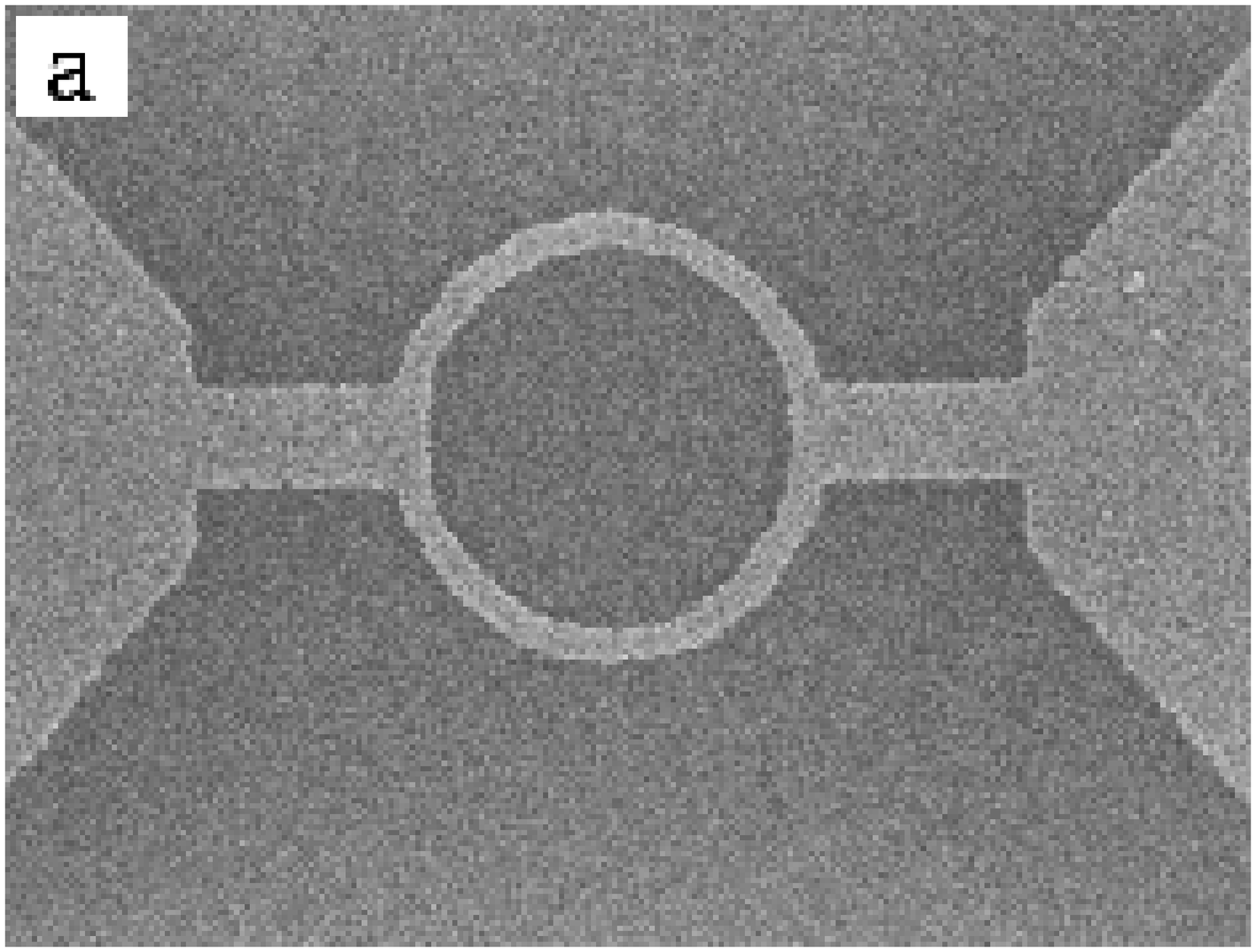,width=0.45\columnwidth} &
         \psfig{file=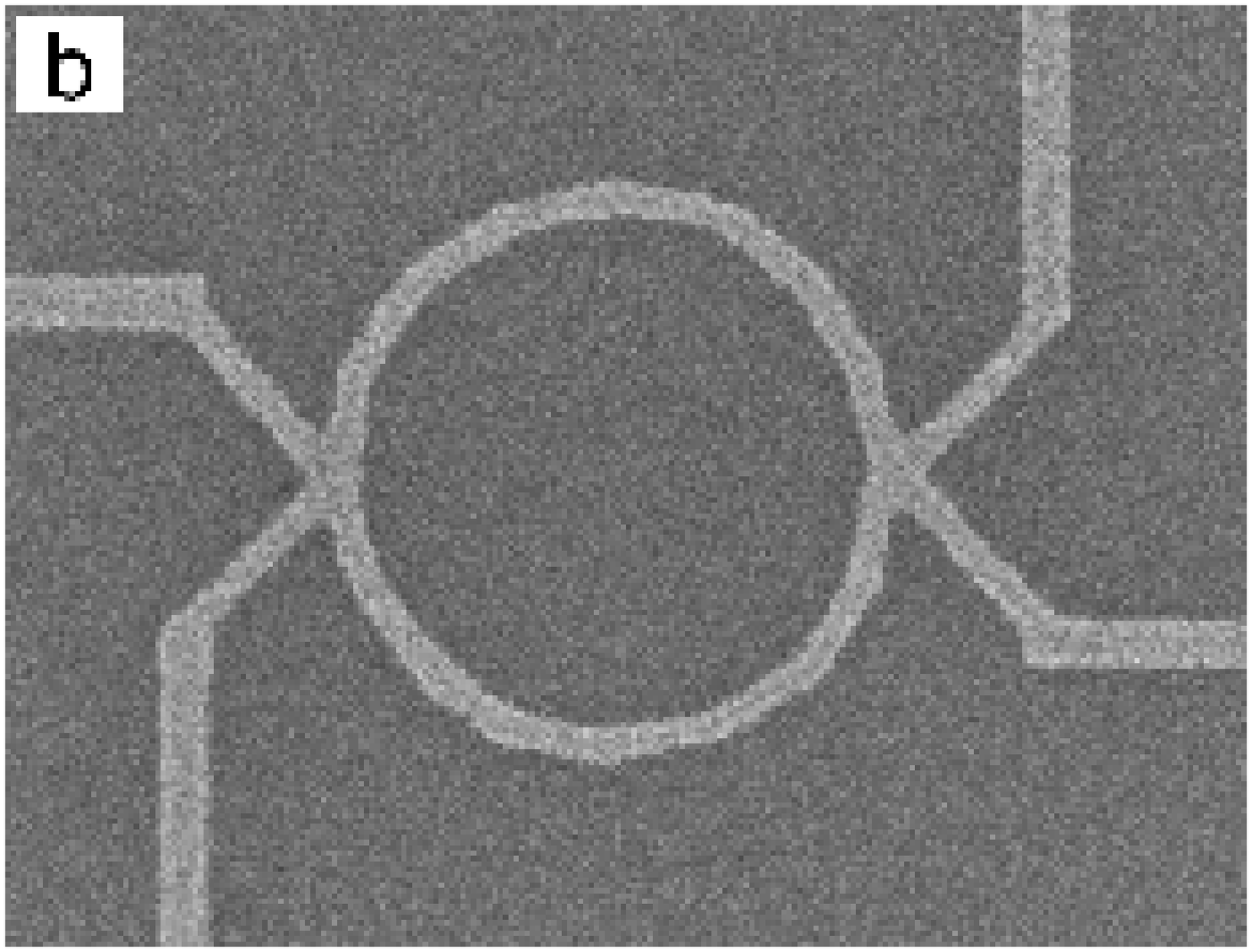,width=0.45\columnwidth} \\[1em]
      \end{tabular}
      \caption{Electron microscope photographs of a ring in a) two-wire
               and b) four-wire configuration.
               The rings have a diameter of 1~$\mu$m and a linewidth of a) 
	       80~nm and b) 60~nm.
	       The thickness of the Cu film is 15~nm.}
      \label{fig1}
   \end{center}
\end{figure}

Figure~\ref{fig1} shows electron-microscope photographs of two rings.
In the two-wire configuration the voltage and current leads branch at a
distance of 4~$\mu$m from the ring (not shown in the photograph) which is
more than the phase-coherence length of $\approx 1~\mu$m.
In the four-wire configuration the bifurcation is directly at the ring.

\section{Results and discussion}

\begin{figure}
   \centerline{\psfig{file=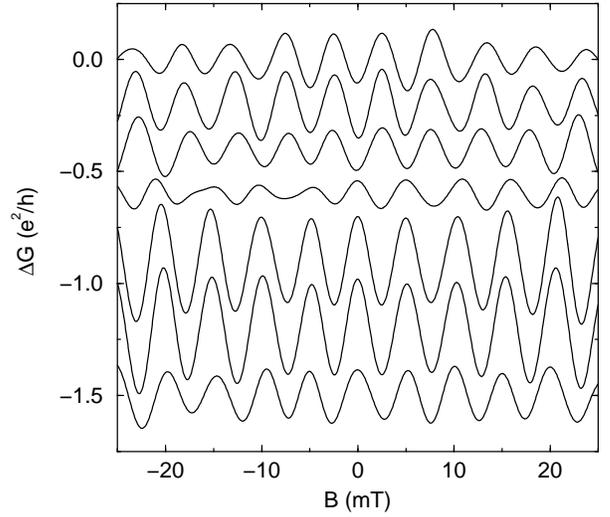,width=0.9\columnwidth}}
   \caption{Conductance $\Delta G$ of a ring in two-wire configuration
            vs. magnetic field $B$ at a temperature $T=90~$mK and currents
	    $I_{DC} = 14.0, 14.4, 14.8, \ldots 16.4~\mu {\rm A}$
            (from top to bottom).
            The data were digitally Fourier filtered.
            Only frequencies corresponding to a period range from 3~mT to 10~mT
	    were taken into account.
            The data are offset vertically for clarity.}
   \label{fig2}
\end{figure}

Figure~\ref{fig2} displays the magnetoconductance of a ring in two-wire
configuration.
The magnetic-field axis was corrected by an offset of 4~mT
(which was the same for all measurements)
due to the hysteresis of the superconducting magnet.
The AB oscillations with a period of 5.2~mT are clearly visible
and it is evident that their amplitude and phase vary with $I_{DC}$.
At $I_{DC}=15.2~\mu$A the amplitude has a minimum and the phase has flipped 
by $\pi$.
In a separate publication~\cite{haussler} we showed that the average
amplitude of the oscillations increases with increasing $I_{DC}$.
For the evaluation of the phase shift we used this effect to measure
oscillations with larger amplitude at large $I_{DC}$ and hence to increase
the signal-to-noise ratio.
Therefore the data presented here were measured at a current of several $\mu$A.
The AB oscillations of a ring in four-wire configuration (not shown) are also
shifted with $I_{DC}$, but they are not symmetric upon reversal of $B$, 
in agreement with theoretical prediction~\cite{buttiker}
and previous experimental observation~\cite{benoit}.

The average phase shift in the whole investigated magnetic-field range is
analyzed quantitatively by calculating the cross-correlation function (CCF)
between the conductance oscillations measured at currents $I_1$ and $I_2$:
\begin{displaymath}
C(I_1,I_2,\Delta B) = \int %\limits_{B_1}^{B_2} 
            \, \Delta G_{I_1}(B) \, \Delta G_{I_2}(B+\Delta B) \, dB
\end{displaymath}
The CCF of two periodic functions with the same period is again a periodic
function. A shift of $\delta B$ between these functions manifests itself
in a shift of the maxima of $C(\Delta B)$ by $\delta B$.

\begin{figure}
   \centerline{\psfig{file=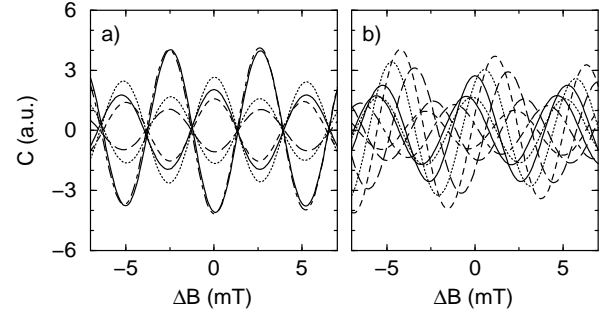,width=0.9\columnwidth}}
   \caption{a) Cross-correlation functions $C(I_1,I_2,\Delta B)$
            for $I_1=14.0~\mu$A and different $I_2=I_{DC}$  vs. $\Delta B$
            for a ring in two-wire configuration.
            The magnetoconductance is displayed in fig.~\ref{fig2} and was
	    evaluated in a magnetic-field range from $-30~$mT to 30~mT.
            b) Cross-correlation functions for a ring in four-wire
	    configuration for comparison.}
   \label{fig3}
\end{figure}

Figure~\ref{fig3}a shows the evaluation of the oscillations displayed in
fig.~\ref{fig2} for a ring in the two-wire configuration.
The CCF's are calculated for $\Delta G(B)$ taken at different $I_2=I_{DC}$,
each time with reference to $\Delta G(B)$ at $I_1=14.0~\mu$A.
It can be seen that the oscillations are either in phase or shifted by
$\delta B = B_{per}/2$.
For comparison the evaluation for a ring in four-wire configuration is
displayed in fig.~\ref{fig3}b.
Here the shift $\delta B$ between the oscillations is arbitrary.

\begin{figure}
   \centerline{\psfig{file=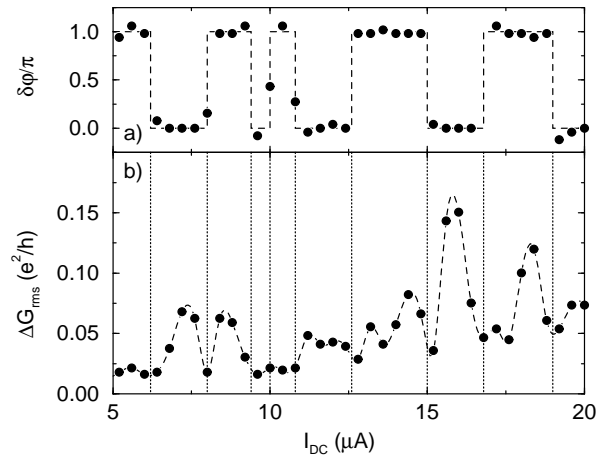,width=0.9\columnwidth}}
   \caption{a) Phase shift referred to $I_{DC}=7.6~\mu$A and
            b) rms-amplitude of the conductance oscillations vs. $I_{DC}$.
            Some of the magnetoconductance traces are displayed in
	    fig.~\ref{fig2}.
            The data were measured at a ring in two-wire configuration and
            evaluated in a magnetic-field range from $-30~$mT to 30~mT.
            The dashed lines are guides to the eye.
            The dotted vertical lines mark the positions of a phase flip
	    by $\pm \pi$.}
   \label{fig4}
\end{figure}

The phase shift of the oscillations $\delta \varphi$ is related to
$\delta B$ by the relation
\mbox{$\delta \varphi = 2 \pi \delta B / B_{per}$}.
The observed shift of the CCF's results in $\delta \varphi = 0$ 
or $\delta \varphi = \pi$ for a two-wire configuration (cf. fig.~\ref{fig4}a)
whereas $\delta \varphi$ is arbitrary for a four-wire configuration.
This behaviour demonstrates convincingly the expected symmetry relations
as the magnetoconductance of a sample in two-wire configuration has to be
symmetric upon reversal of $B$.
This restricts the phase $\varphi$ in eq.~(\ref{eq1}) to either 0 or
$\pi$ whereas there is no constraint for a four-wire configuration.
A quantitative analysis performed over a wide range of $I_{DC}$ revealed
that the typical current scale for phase flips corresponds to the
Thouless energy $E_c$, in agreement with eq.~(\ref{eq1}).

From a simple point of view the AB oscillations arise from the interference
of an electron which splits into two partial waves at one side of the ring.
These waves traverse the ring in opposite arms and recombine at the other side.
The interference is determined by the phase difference of the wave functions.
From this simple argument the phase of the wave functions should be the same
at $B=0$ provided that the arms have equal lengths.
Hence at $B=0$ the interference should be constructive and the conductance
have a maximum.

However, the phase of the electronic wave function also depends on the
configuration of the scattering centres.
For this reason the phase will usually not be the same in both arms of the
ring,
so that on the one hand at $B=0$ an arbitrary interference between the partial
waves might be possible.
Indeed it was demonstrated that the AB oscillations average to zero when
the measurements are taken of a series of rings~\cite{umbach}.
On the other hand the symmetry relations for the magnetoconductance require
that for a ring in a two-wire configuration the conductance at $B=0$ has
either a maximum or a minimum.
This means that the interference has to be either constructive or destructive.

The key to the resolution of these seemingly contradictory statements
is that the above argument considering only two wave functions is too simple.
For the correct calculation of the conductance all transmitted and reflected
partial waves have to be taken into account.
However, an interesting question arises when we vary the phases of the
electronic wave functions continuously by a modification of $I_{DC}$:
what is the reaction of the system to this continuous variation of the
phases of the wave functions under the condition that the macroscopically
observable phase of the conduction oscillations is quantized?

For this purpose the phase and the amplitude of the oscillations are
displayed in a common graph in fig.~\ref{fig4}.
The data were extracted from a series of magnetoconductance traces
(some of them are shown in fig.~\ref{fig2})
measured on a ring in two-wire configuration.
It can be seen that both quantities are a function of $I_{DC}$.
There is a definite correlation between the variation of the amplitude
and the phase:
at each phase flip by $\pi$ there is a minimum of the oscillation amplitude.
Hence no abrupt flip of the conductance oscillations occurs,
but every variation of the phase by $\pm \pi$ is accompanied by a continuous
decrease and subsequent increase of the oscillation amplitude.
Superimposed on the strong fluctuations of the oscillation amplitude
an increase of $\Delta G_{rms}$ is observed.
Averaging over a larger field interval than just $[-30~{\rm mT}, 30~{\rm mT}]$
yields $\Delta G_{rms} \sim \sqrt{I_{DC}}$,
in agreement with the prediction for the conductance fluctuations in singly
connected mesoscopic samples~\cite{haussler}.

In conclusion, although the microscopic phase of the electronic wave function
is varied continuously by a modification of $I_{DC}$
the macroscopic phase of the conductance oscillations varies in a quantized
manner.
However, there is no abrupt change in the magnetoconductance.
Rather, the macroscopic phase flip is accommodated by a rearrangement of the
individual electron phases to produce an interference pattern leading to
a minimum of the oscillation amplitude.
This continuous variation of the interference on the microscopic level
is directly visible (cf. fig.~\ref{fig2}) as a slight shift of the oscillation
frequency whenever a phase flip occurs.
In a recent experiment on semiconductor rings in two-wire configuration
with a few electron transmission channels the phase of the oscillations
could be modified by variation of a gate voltage~\cite{hansen}.
A phase flip by $\pi$ of the fundamental oscillation was accompanied
by the occurrence of a dominating higher harmonic with $h/2e$ periodicity.
In our experiment on diffusive rings with many electron transmission channels
we observe oscillations with only the fundamental $h/e$ periodicity
whose suppression at macroscopic phase flips indicates the rearrangement
at the microscopic level.

\section{Acknowledgements}

We thank P. Pfundstein for operation of the electron microscope.
This work was supported by the Deutsche Forschungsgemeinschaft through SFB 195.

\end{document}